\documentclass[twocolumn,english,prl, merge, semicolon]{revtex4}
\pdfoutput=1
\usepackage[T1]{fontenc}
\usepackage[utf8]{inputenc}
\usepackage{amsmath}
\usepackage{amssymb}
\usepackage{graphicx}

\makeatletter
\@ifundefined{textcolor}{}
{%
 \definecolor{BLACK}{gray}{0}
 \definecolor{WHITE}{gray}{1}
 \definecolor{RED}{rgb}{1,0,0}
 \definecolor{GREEN}{rgb}{0,1,0}
 \definecolor{BLUE}{rgb}{0,0,1}
 \definecolor{CYAN}{cmyk}{1,0,0,0}
 \definecolor{MAGENTA}{cmyk}{0,1,0,0}
 \definecolor{YELLOW}{cmyk}{0,0,1,0}
 }

\usepackage{hyperref}
\hypersetup{colorlinks=true}

\makeatother

\usepackage{babel}
\begin{document}

\title{Spatial Analogue of Quantum Spin Dynamics via Spin-Orbit Interaction}

\author{Vanita Srinivasa}

\email{vas9@pitt.edu}

\address{Department of Physics and Astronomy, University of Pittsburgh, Pittsburgh,
Pennsylvania 15260, USA}

\author{Jeremy Levy}

\email{jlevy@pitt.edu}

\address{Department of Physics and Astronomy, University of Pittsburgh, Pittsburgh,
Pennsylvania 15260, USA}
\begin{abstract}
We map electron spin dynamics from time to space in quantum wires
with spatially uniform and oscillating Rashba spin-orbit coupling.
The presence of the spin-orbit interaction introduces pseudo-Zeeman
couplings of the electron spins to effective magnetic fields. We show
that by periodically modulating the spin-orbit coupling along the
quantum wire axis, it is possible to create the spatial analogue of
spin resonance, without the need for any real magnetic fields. The
mapping of time-dependent operations onto a spatial axis suggests
a new mode for quantum information processing in which gate operations
are encoded into the band structure of the material. We describe a
realization of such materials within nanowires at the interface of
LaAlO$_{3}$/SrTiO$_{3}$ heterostructures. 
\end{abstract}
\maketitle
Quantum dynamics lies at the heart of modern physics. While the evolution
of a quantum-mechanical system is governed by the time-dependent Schrödinger
equation (TDSE), the mapping of this evolution from time to space
has given rise to fundamental advances through the development of
powerful theoretical techniques. One familiar example is provided
by Feynman's path integral method \citep{Feynman1948}, in which a
description equivalent to the TDSE is obtained by recasting time evolution
in terms of space-time paths. This approach finds applicability in
a wide variety of contexts ranging from relativistic quantum mechanics
to the braiding of quasiparticles that forms the basis for topological
quantum computation \citep{Nayak2008}. World lines associated with
quasiparticle braiding in two dimensions can be mapped to flux lines
in three dimensions \citep{Nelson1988} if time itself is treated
as a spatial coordinate by analytical continuation to imaginary time.
In general, combining imaginary time with $d$ spatial dimensions
defines a ($d+1$)-dimensional Euclidean space that provides a correspondence
between quantum field theory and statistical mechanics \citep{Tsvelik1995}.
Similarly, a connection between quantum field theory and quantum gravitational
theory emerges through holographic mapping of a ($d+1$)-dimensional
combined space-time description to a $d$-dimensional one \citep{Susskind1995}. 

In the present work, we explore a mapping of spin dynamics from time
to space, motivated by its potentially fundamental relevance to methods
for coherent spin manipulation in solid-state systems \citep{Hanson2008}
and physical realizations of spin-based quantum computing \citep{DiVincenzo2000,Burkard2000}.
Addressing these challenges involves harnessing the interactions of
spin with spatial as well as external degrees of freedom. We show
here that the coupling between spin and space can in fact be used
to map the spin evolution from a temporal axis to a spatial one. Conceptually,
instead of the TDSE (for $\hbar=1$), 
\begin{equation}
i\partial_{t}U\left(t\right)=H\left(t\right)U\left(t\right),\label{eq:tdse}
\end{equation}
the evolution is governed by a spatial analogue, which we write as
\begin{equation}
i\partial_{y}U\left(y\right)=K\left(y\right)U\left(y\right).\label{eq:sdse}
\end{equation}
The coordinate-dependent {}``quasimomentum operator'' $K\left(y\right)$
in Eq. \eqref{eq:sdse} plays a role similar to a time-varying Hamiltonian.
The time-to-space mapping of spin dynamics then entails identifying
a form for $K\left(y\right)$ that generates a unitary transformation
$U\left(y\right)$ describing a simultaneous spatial translation and
spin rotation. This identification can be made by considering the
spatial analogues of time-dependent Hamiltonians $H\left(t\right)$
which generate spin dynamics.

Electron spin resonance (ESR) enables three-dimensional dynamical
manipulation of single electron spins and therefore plays a central
role in their promise as natural candidates for solid-state quantum
bits (qubits) \citep{Loss1998,Hanson2007}. While the Zeeman interaction
describing the coupling of electron spin to external magnetic fields
is typically used to carry out ESR, applying the local magnetic fields
required for selectively addressing individual electron spin qubits
is challenging in practice. The implementation of ESR using electric
fields has therefore been widely investigated, leading to methods
such as g-tensor modulation resonance (g-TMR) \citep{Kato2003}, electric
dipole spin resonance (EDSR) \citep{Rashba2003,Kato2004,Duckheim2006,Golovach2006,Nowack2007},
and ballistic spin resonance (BSR) \citep{Frolov2009}. 

A basic resource in both EDSR and BSR is the spin-orbit interaction,
which couples the spin of an electron to its spatial motion in physical
systems lacking inversion symmetry \citep{Winkler2003}. This coupling
has the general form $H_{so}\propto\left(\vec{\nabla}V\times\vec{k}\right)\cdot\vec{\sigma}$
and describes the interaction of the electron spin {[}represented
by the vector of Pauli operators $\vec{\sigma}=\left(\sigma_{x},\sigma_{y},\sigma_{z}\right)${]}
with an \emph{effective} magnetic field $\vec{B}_{eff}\propto\left(\vec{\nabla}V\times\vec{k}\right)$
that allows for the manipulation of spins through both an electric
field $\vec{E}$ via $\vec{\nabla}V\propto\vec{E}$ and the electronic
momentum $\vec{\hbar k}$. A time-dependent oscillation of the spin-orbit
interaction, generated by an external driving voltage in EDSR and
internal electron dynamics in BSR, replaces the oscillating magnetic
field required for ESR in these methods. The effective magnetic field
due to spin-orbit coupling is also the key ingredient in several proposals
for single-qubit gates \citep{Popescu2004,Stepanenko2004,Flindt2006,Gong2007,Golovach2010,Nadj-Perge2010}.

Here, we describe a mechanism for spin resonance that relies on \emph{spatially}-varying
spin-orbit coupling and the associated effective magnetic field, without
employing any real magnetic fields. This method maps the spin evolution
from time to space and is therefore not subject to the typical time-dependent
constraints imposed in quantum computing for the purpose of preserving
coherence \citep{DiVincenzo2000,Burkard2000}. We show theoretically
that this {}``spin spatial resonance'' can be achieved by creating
a superlattice within a quantum wire via periodically-modulated asymmetry
in the lateral confinement potential. Rather than being controlled
by time-dependent external fields, spin spatial resonance is built
into the spin-dependent band structure \citep{Demikhovskii2007,Smirnov2007}
of the superlattice. Segments of this “designer quantum material”
having fixed lengths can be used to apply spatial “pulses” that execute
single-qubit gate operations on the spins of electrons which travel
through the wires. Because the system is one-dimensional and gate
operations are determined only by the spatial coordinate of an electron,
these single-qubit gates are intrinsically “fault-tolerant” with respect
to backscattering, as we discuss below. 

\begin{figure}
\includegraphics[width=3.375in]{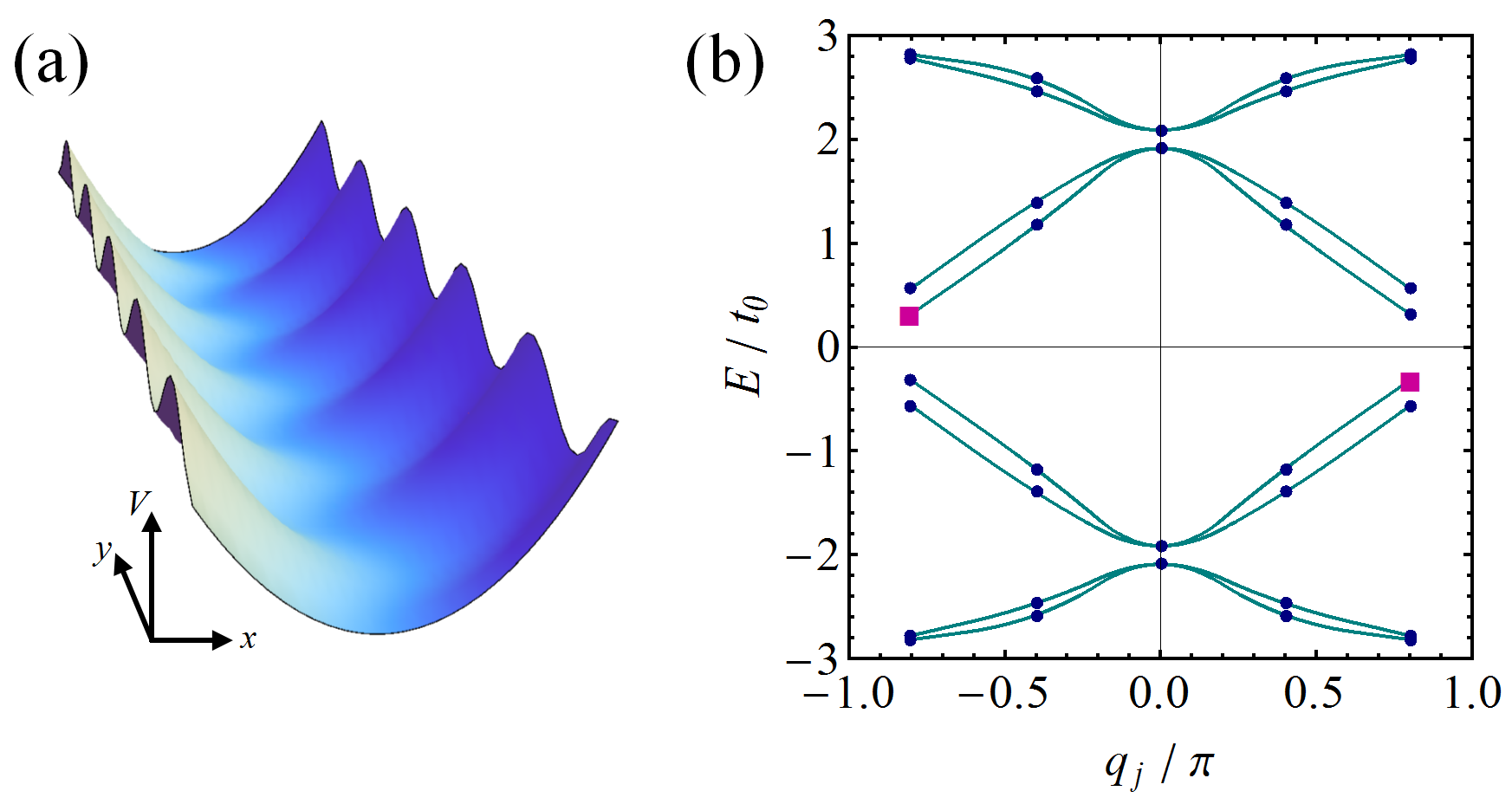}\caption{Spin-orbit superlattice quantum wire. (a) Schematic of a quantum wire
with periodically-varying lateral confinement asymmetry which gives
rise to a spin-orbit superlattice within the wire. (b) Spectrum as
a function of electron quasimomentum $q_{j}$ in the presence of both
spatially uniform and oscillating Rashba spin-orbit coupling in perpendicular
directions for $N=20,$ $m=4,$ $t_{so}^{unif}/t_{0}=1,$ and $t_{so}^{osc}/t_{0}=0.125$.
Filled squares indicate the eigenstates used to form the wavepacket
state $\left|\Psi_{wp}\right\rangle $ discussed in the text. Lines
are guides to the eye.}
\end{figure}

In order to show how ESR can be mapped to space using the spin-orbit
interaction alone, we first describe a {}``spin-orbit superlattice
quantum wire,'' i.e., a one-dimensional system with built-in spatially
uniform \citep{Governale2004,Smirnov2007} and oscillating \citep{Wang2004,Demikhovskii2007,Japaridze2009}
spin-orbit coupling in perpendicular directions {[}Fig. 1(a){]}. For
specificity, we choose the coupling to be of the Rashba form \citep{Bychkov1984},
although it may be possible to realize the mapping presented here
with other types of spin-orbit interaction. We assume a zero-temperature
independent-electron description and represent the quantum wire confining
the electronic motion to one spatial dimension using a single-band
tight-binding model. In addition to a term $H_{0}$ describing the
kinetic hopping of the electron along the wire, a tight-binding Hamiltonian
spectrally equivalent to the Rashba spin-orbit interaction via a spin-dependent
hopping can be written by discretizing the interaction on a lattice
\citep{Mireles2001,Metalidis2007thesis}. We extend this tight-binding
model in order to incorporate spatially-varying Rashba spin-orbit
coupling. The full Hamiltonian for a wire represented by $N$ sites
with lattice constant $a$ is given by 
\begin{equation}
H=H_{0}+H_{so}^{unif}+H_{so}^{osc},\label{eq:h}
\end{equation}
where
\begin{eqnarray}
H_{0} & = & -t_{0}\sum\limits _{n,\sigma}\left(c_{n+1,\sigma}^{\dagger}c_{n,\sigma}+\textrm{H.c.}\right),\label{eq:h0tb}\\
H_{so}^{unif} & = & -t_{so}^{unif}\sum\limits _{n,\sigma,\sigma'}\left[c_{n+1,\sigma'}^{\dagger}\left(i\sigma_{x}\right)_{\sigma'\sigma}c_{n,\sigma}\right.\nonumber \\
 &  & \left.\qquad\qquad\qquad+\:\textrm{H.c.}\right],\label{eq:hsouniftb}\\
H_{so}^{osc} & = & -t_{so}^{osc}\sum\limits _{n,\sigma,\sigma'}\varphi_{n}\left[c_{n+1,\sigma'}^{\dagger}\left(i\sigma_{z}\right)_{\sigma'\sigma}c_{n,\sigma}\right.\nonumber \\
 &  & \left.\qquad\qquad\qquad\;\;+\:\textrm{H.c.}\right].\label{eq:hsoosctb}
\end{eqnarray}
Here, the operators $c_{n,\sigma}^{\dagger}$ and $c_{n,\sigma}$
create and annihilate, respectively, an electron at site $n$ with
spin $\sigma,$ and $t_{0}=\hbar^{2}/2m^{*}a^{2},$ where $m^{*}$
is the effective mass of the electron. The spin-dependent hopping
amplitude $t_{so}^{unif}=\alpha_{\bot}/2a$ describes spatially uniform
Rashba spin-orbit coupling of strength $\alpha_{\bot}$ generated
by the potential gradient perpendicular to the plane containing the
wire (defined to be the $x-y$ plane). Spatially-varying Rashba spin-orbit
coupling with amplitude $\alpha_{\parallel}$ due to the potential
asymmetry in the $x-y$ plane but perpendicular to the propagation
direction ($y$ axis) of the wire is incorporated via a hopping described
by an amplitude $t_{so}^{osc}=\alpha_{\parallel}/2a$ together with
a site-dependent factor $\varphi_{n}$. The summations in Eqs. \eqref{eq:h0tb}-\eqref{eq:hsoosctb}
run over $n=1,\ldots,\, N$ and $\sigma,\sigma'=\uparrow,\downarrow$.
Here, we assume periodic boundary conditions so that $n\pm N\equiv n.$
Note that we choose the uniform {[}Eq. \eqref{eq:hsouniftb}{]} and
oscillating {[}Eq. \eqref{eq:hsoosctb}{]} effective magnetic fields
to lie along the $x$ and $z$ axes, respectively. In the present
work, we therefore define $\left|\sigma\right\rangle \equiv\left|\sigma\right\rangle _{x}$;
i.e., we choose the basis $\left\{ \left|\uparrow\right\rangle _{x},\left|\downarrow\right\rangle _{x}\right\} $
associated with the components of spin along the $x$ axis in order
to describe the spin orientation with respect to the effective magnetic
field due to the uniform spin-orbit interaction {[}Eq. \eqref{eq:hsouniftb}{]}. 

In the presence of only uniform spin-orbit coupling ($t_{so}^{osc}=0$),
the energy spectrum for the Hamiltonian consists of two tight-binding
bands. These bands are shifted in opposite directions along the quasimomentum
axis with respect to the energy band for $H_{0}$ due to the momentum
dependence of the effective magnetic field associated with the spin-orbit
interaction. If the modulated Rashba spin-orbit coupling term in Eq.
\eqref{eq:hsoosctb} is included in the Hamiltonian ($t_{so}^{osc}\neq0$),
the translational symmetry of the system is reduced. To map ESR spatially,
we introduce periodically-varying spin-orbit coupling via $\varphi_{n}=\cos\left(2\pi n/m\right).$
The Hamiltonian in Eq. \eqref{eq:h} then retains a periodicity over
$m$ lattice sites. In this case, the energy eigenstates can be characterized
by the eigenvalues of an operator $D_{m},$ which is defined to be
the discrete translation operator over $m$ sites and for which $\left[H,\: D_{m}\right]=0$.
The eigenvalues of $D_{m}$ are of the form $e^{iq_{j}}$, where the
associated dimensionless quasimomenta are $q_{j}\equiv2\pi j/N',$
with $N'\equiv N/m$ and $j$ an integer such that $-N'/2\leq j<N'/2.$
Letting $n=mn'+l,$ we express the Hamiltonian in terms of the operators
$\tilde{c}_{j,l,\sigma}\equiv\frac{1}{\sqrt{N'}}\sum_{n'=0}^{N'-1}e^{-iq_{j}\left(mn'+l\right)}c_{mn'+l,\sigma}$
and $\tilde{c}_{j,l,\sigma}^{\dagger}.$ Subsequently transforming
to the representation given by $d_{j,p,\sigma}\equiv\frac{1}{\sqrt{m}}\sum_{l=1}^{m}e^{-iQ_{p}l}\tilde{c}_{j,l,\sigma}$
and $d_{j,p,\sigma}^{\dagger},$ where $Q_{p}\equiv2\pi p/m$ with
$p$ an integer such that $-m/2\leq p<m/2,$ leads to the following
equivalent forms for Eqs. \eqref{eq:h0tb}-\eqref{eq:hsoosctb}: 
\begin{eqnarray}
H_{0} & = & -2t_{0}\sum\limits _{j,p,\sigma}\cos\left(q_{j}+Q_{p}\right)d_{j,p,\sigma}^{\dagger}d_{j,p,\sigma},\label{eq:h0k}\\
H_{so}^{unif} & = & -2t_{so}^{unif}\sum\limits _{j,p,\sigma}\sigma\sin\left(q_{j}+Q_{p}\right)d_{j,p,\sigma}^{\dagger}d_{j,p,\sigma},\label{eq:hsounifk}\\
H_{so}^{osc} & = & -t_{so}^{osc}\sum\limits _{j,p,\sigma,\sigma'}\left(1-\delta_{\sigma',\sigma}\right)\sin\left(q_{j}+Q_{p}+\frac{\pi}{m}\right)\nonumber \\
 &  & \times\left(e^{-i\pi/m}d_{j,p+1,\sigma'}^{\dagger}d_{j,p,\sigma}+\textrm{H.c.}\right).\label{eq:hsoosck}
\end{eqnarray}
From the above expressions, it is evident that the kinetic hopping
term $H_{0}$ and the uniform spin-orbit coupling term $H_{so}^{unif}$
are diagonal in the basis corresponding to the operators $d_{j,p,\sigma}^{\dagger}$
and $d_{j,p,\sigma}$, where $j$ indicates the quasimomentum $q_{j},$
$p$ is a band index, and $\sigma$ represents the spin component
along the $x$ axis {[}defined to be 1 (-1) for $\uparrow\left(\downarrow\right)$
when used to explicitly represent the eigenvalue of $\sigma_{x}$,
as in Eq. \eqref{eq:hsounifk}{]}. The expression for $H_{so}^{osc}$
in this basis implies that the periodically-modulated spin-orbit interaction
couples {}``adjacent'' ($\Delta p=1$) bands having opposite spin
($\sigma'\neq\sigma)$. The symmetry associated with this coupling
of basis states gives rise to $m$-dimensional representations of
the Hamiltonian. To obtain the full spectrum for Eq. \eqref{eq:h},
we numerically diagonalize the Hamiltonian matrix within each of the
two $m$-dimensional subspaces associated with each value of $q_{j}.$
As in standard ESR, we treat the oscillating effective magnetic field
as a perturbation relative to the uniform effective field and choose
$t_{so}^{osc}/t_{0}\ll t_{so}^{unif}/t_{0}$. Figure 1(b) shows the
spectrum as a function of $q_{j}$ for $N=20,$ $m=4,$ $t_{so}^{unif}/t_{0}=1,$
and $t_{so}^{osc}/t_{0}=0.125,$ illustrating the coupling between
the bands due to $H_{so}^{osc}.$

\begin{figure}
\includegraphics[width=3.375in]{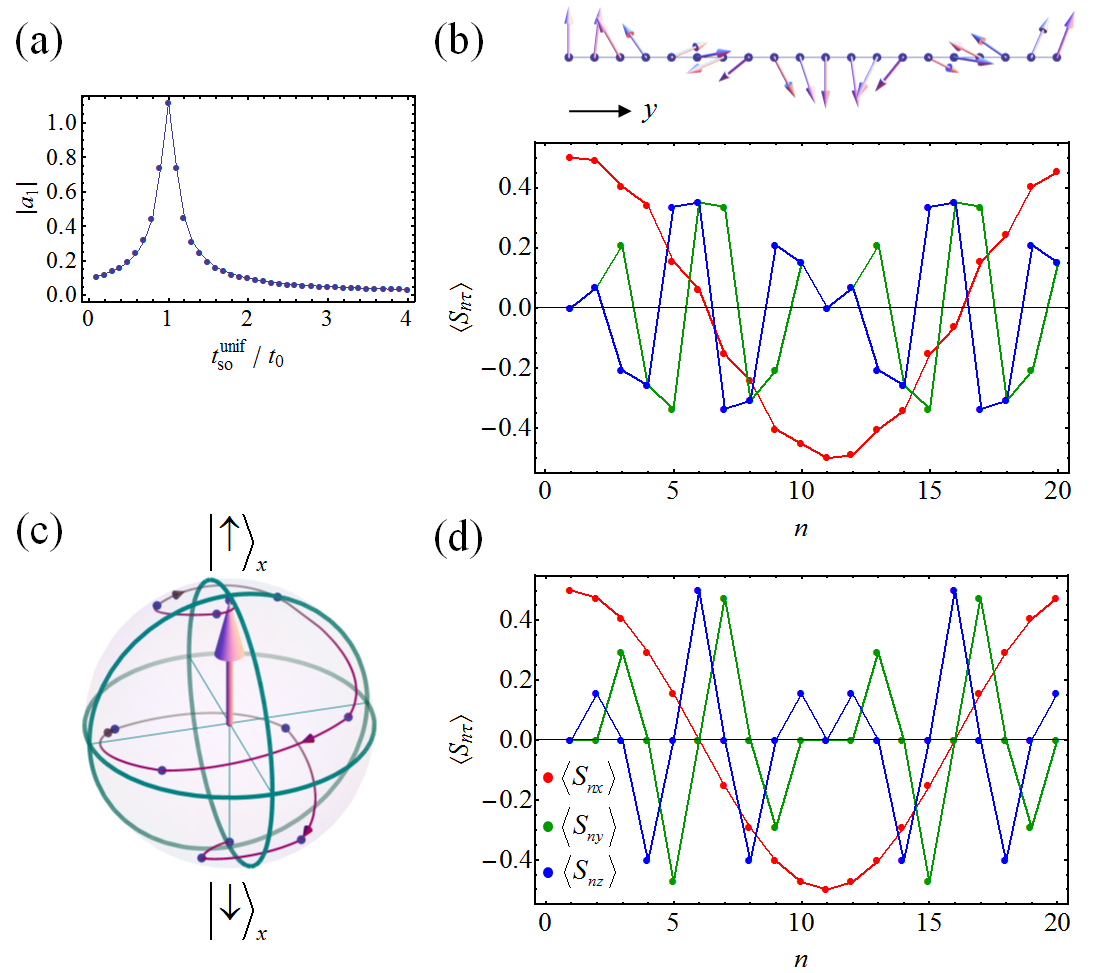}\caption{Spin spatial resonance. (a) Amplitude for the lowest nonzero ($l=1$)
Fourier mode of the oscillation in $\left\langle S_{nx}\right\rangle $
as a function of $t_{so}^{unif}/t_{0}$, showing a peak at  $t_{so}^{unif}/t_{0}=1$.
(b) Spin polarization (in units of $\hbar$) for the state $\left|\Psi_{wp}\right\rangle $
at $t_{so}^{unif}/t_{0}=1$ as a function of spatial coordinate along
the propagation direction ($y$ axis) of the wire. (c) Bloch-sphere
representation of the spin polarization in (b). As a function of the
spatial coordinate $y$ along the wire, the spin polarization follows
a spiraling trajectory typical of spin resonance. Points indicated
along the trajectory correspond to lattice sites. (d) Spin polarization
components as a function of spatial coordinate for the analytical
continuum model based on Eqs. \eqref{eq:sdse} and \eqref{eq:Ky}.}
\end{figure}

To demonstrate that signatures of spin resonance exist in the band
structure of a spin-orbit superlattice quantum wire, we form a superposition
of two energy eigenstates with equal and opposite quasimomenta $q_{j}$
{[}indicated by squares in Fig. 1(b){]}. Expressing these eigenstates
using the notation $\left|j,\nu\right\rangle ,$ where $\nu=0,1,\ldots7$
represents a combined spin-orbital index that is defined to increase
with increasing energy $E$, we define a {}``wavepacket'' $\left|\Psi_{wp}\right\rangle \equiv\frac{1}{\sqrt{2}}\left(\left|-2,\,4\right\rangle +e^{-i\phi}\left|2,\,3\right\rangle \right).$
Note that the variation of $\phi$, which changes the relative phase
between the two eigenstates, is equivalent to the time evolution of
$\left|\Psi_{wp}\right\rangle $ and simply results in a shift of
the phase of the oscillation in the spin polarization. We therefore
fix $\phi$ and calculate the spin polarization as a function of site
$n$ for the wavepacket by determining the expectation values $\left\langle S_{n\tau}\right\rangle \equiv\left\langle \Psi_{wp}\mid S_{n\tau}\mid\Psi_{wp}\right\rangle $,
where $S_{n\tau}\equiv\left|n\right\rangle \left\langle n\right|\otimes\frac{\hbar}{2}\sigma_{\tau}$
is the spin operator at site $n$ and $\tau=x,y,z.$ In the following,
we let $\hbar=1$. We use the discrete Fourier transform $F_{d},$
defined by $\left\langle l\mid F_{d}\mid n\right\rangle \equiv\left(1/\sqrt{N}\right)e^{-2\pi il(n-1)/N}$,
to calculate the distribution of the Fourier modes in $\left\langle S_{nx}\right\rangle $
as a function of $t_{so}^{unif}/t_{0}$. The absolute value of the
amplitude of the $l=1$ Fourier mode, $\left|a_{1}\right|$ {[}Fig.
2(a){]}, has a peak for $t_{so}^{unif}/t_{0}=1$. A calculation of
the spin polarization components $\left(\left\langle S_{nx}\right\rangle ,\left\langle S_{ny}\right\rangle ,\left\langle S_{nz}\right\rangle \right)$
for the wavepacket corresponding to this peak {[}Fig. 2(b){]} reveals
that, while $\left\langle S_{ny}\right\rangle $ and $\left\langle S_{nz}\right\rangle $
vary rapidly over space, $\left\langle S_{nx}\right\rangle $ exhibits
a more gradual oscillation with one full cycle over the length of
the wire. The corresponding spatial dependence of the spin polarization
vector is illustrated above the plot of the spin polarization components
in Fig. 2(b). The Bloch-sphere evolution of the spin vector {[}Fig.
2(c){]} follows a spiraling path typical of ESR. Here, however, the
evolution of the spin occurs with respect to a spatial coordinate
- the distance along the $y$ axis. Calculations performed for a wavepacket
constructed from the two ground states $\left|\pm2,0\right\rangle $
reveal similar results, with a peak in $\left|a_{1}\right|$ at $t_{so}^{unif}/t_{0}=1$. 

The above analysis based on the tight-binding model defined in Eqs.
\eqref{eq:h}-\eqref{eq:hsoosctb} is well described by an analytical
model obtained from a mapping of the standard electron spin resonance
formalism for a two-state system from time to space via Eq. \eqref{eq:sdse}.
Based on the continuum version of Eq. \eqref{eq:h}, we choose the
form 
\begin{equation}
K\left(y\right)\equiv\frac{k_{0}}{2}\sigma_{x}-k_{1}\cos\left(ky\right)\sigma_{z}\label{eq:Ky}
\end{equation}
for the quasimomentum operator. Here, $k\equiv2\pi/\lambda$ is the
spatial frequency associated with oscillating spin-orbit coupling
of wavelength $\lambda$. Eq. \eqref{eq:Ky} has a form analogous
to a spin resonance Hamiltonian, with time replaced by the spatial
coordinate $y$ along the wire and temporal frequencies replaced by
their spatial counterparts. The spin polarization components can be
written in terms of the matrix elements of the solution $U\left(y\right),$
which can be determined using spatial analogues of the interaction
picture, the rotating wave approximation, and a transformation to
a rotating frame. Choosing $\lambda=4a,$ $k_{0}=\left(\pi/2a\right)\left(t_{so}^{unif}/t_{0}\right),$
and $k_{1}=\left(4\pi/5a\right)\left(t_{so}^{osc}/t_{0}\right)$ in
order to make a correspondence with the wavepacket state used in the
tight-binding calculation described above, we let $y=na$ and evaluate
the spin polarization as a function of $n$. The case $t_{so}^{unif}/t_{0}=1,$
$t_{so}^{osc}/t_{0}=0.125$, which corresponds to the spatial resonance
condition $k=k_{0}$ and the fundamental spatial frequency $k_{1}=\pi/10a,$
is shown in Fig. 2(d). This analytical result for the spin polarization
components possesses qualitative features similar to the numerical
result based on diagonalization of the tight-binding model for the
same values of $t_{so}^{unif}/t_{0}$ and $t_{so}^{osc}/t_{0}$ {[}Fig.
2(b){]}. The distortion of $\left\langle S_{nx}\right\rangle $ in
Fig. 2(b) relative to the smooth sinusoidal variation in the continuum
analytical model of Fig. 2(d) is due to the fact that the subspace
of $\left|j,\, p,\,\sigma\right\rangle $ basis states for the tight-binding
calculation consists of $m=4$ rather than two states, all of which
contribute to $\left|\Psi_{wp}\right\rangle $. In addition, the finite
size of the system used in the tight-binding calculation results in
a reduced amplitude of variation in $\left\langle S_{ny}\right\rangle $
and $\left\langle S_{nz}\right\rangle $ compared to the analytical
result of Fig. 2(d). The analytically-obtained signatures of ESR are
nevertheless evident in the tight-binding results for the spin-orbit
superlattice quantum wire, demonstrating that it is indeed possible
to achieve ESR entirely from spin-orbit coupling and map spin resonance
from time to space. 

We now describe a possible physical implementation of a spin-orbit
superlattice quantum wire at the interface of a LaAlO$_{3}$/SrTiO$_{3}$
heterostructure. In this system, a local voltage-induced metal-insulator
transition has been used to demonstrate both the fabrication of nanowires
with widths $\sim2$ nm \citep{Cen2008,*Cen2009} and the incorporation
of highly asymmetric potential profiles along the nanowires \citep{Bogorin2010}.
In principle, the same method can be used to create nanowires with
built-in lateral confinement asymmetry, and periodic variation of
the asymmetry of the applied pulse along the wire can give rise to
spatially oscillating spin-orbit coupling. The corresponding oscillating
effective magnetic field will be oriented perpendicular to the plane
containing the wire (i.e., along the $z$ axis). Together with an
orthogonal effective field due to uniform spin-orbit coupling at the
interface \citep{BenShalom2010,*Caviglia2010}, this would allow for
the creation of a spin-orbit superlattice quantum wire. The resonance
condition $t_{so}^{unif}/t_{0}=1$ can be used to estimate the length
and spatial period of the superlattice. Using the definitions of $t_{0}$
and $t_{so}^{unif}$, we find $a=\hbar^{2}/m^{*}\alpha_{\perp}$.
With $\alpha_{\perp}=8\times10^{-12}\textrm{ eV}\cdot\textrm{m}$
and $m^{*}=1.1m_{e},$ where $m_{e}$ is the free electron mass, $a=300\textrm{ nm.}$
For $m=4,$ this corresponds to $\lambda=ma=4a\sim1\textrm{ \ensuremath{\mu}m}$
for the oscillating spin-orbit coupling {[}Eq. \eqref{eq:hsoosctb}{]}.

The spatial mapping of ESR using only spin-orbit coupling implies
that single-qubit gates can also be mapped to space. These gates are
built into the spin-dependent band structure of the superlattice.
Because the spin polarization is determined by the distance the electron
travels along the wire, segments of spin-orbit superlattice quantum
wires having fixed lengths can be thought of as spatial {}``pulses''
applied to an electron which traverses them. As one application of
this idea, we have found that a spin-echo pulse sequence \citep{Vandersypen2005}
can be mapped from time to space; details of this calculation will
be given in a longer paper. A degree of robustness to spin qubit gate
errors caused by backscattering of the electrons exists by virtue
of the fact that the system is one-dimensional, so that any change
in spin polarization due to backscattering can be undone if the electron
again scatters into its original propagation direction.

While ESR, quantum gates, and pulse sequences are typically regarded
as time-dependent processes, the mapping of these building blocks
for spin manipulation onto a \emph{spatial} axis as described in the
present work suggests a new paradigm for quantum information processing.
In this framework, a spin-orbit superlattice quantum wire represents
a designer quantum material with a single-qubit gate encoded into
its band structure. Generalization of the ideas presented here to
two or more qubits would pave the way for achieving universal quantum
computing via spatial encoding of quantum dynamics. 
\begin{acknowledgments}
We thank J. M. Taylor, G. Burkard, and S. Smirnov for helpful discussions.
This work was supported by NDSEG (V. S.), an Andrew Mellon Fellowship
(V. S.), NSF (DMR-0704022) (J. L.), and ARO MURI (W911NF-08-1-0317)
(J. L.).\bibliographystyle{apsrev}
\bibliography{SOESR}
\end{acknowledgments}

\end{document}